\documentclass[11pt]{article}

\usepackage[preprint]{acl}

\usepackage{times}
\usepackage{latexsym}

\usepackage[T1]{fontenc}

\usepackage[utf8]{inputenc}

\usepackage{microtype}

\usepackage{inconsolata}

\usepackage{graphicx}
\usepackage{booktabs}
\usepackage{array}
\usepackage{amsmath}
\usepackage{amssymb}
\usepackage{pifont}
\usepackage{xcolor}
\definecolor{cmarkgreen}{HTML}{009E73}  
\definecolor{xmarkred}{HTML}{D55E00}    
\newcommand{\cmark}{\textcolor{cmarkgreen}{\ding{51}}}
\newcommand{\xmark}{\textcolor{xmarkred}{\ding{55}}}

\title{Probing Neural Audio Codecs with \\Speaker Traits and Speech Characteristics}

\author{
  \textbf{Jihwan Lee\textsuperscript{1}},
  \textbf{Tiantian Feng\textsuperscript{1}},
    \textbf{Minu Kim\textsuperscript{1}},
  \textbf{Junhyeok Lee\textsuperscript{2}},
    \textbf{Sean Foley\textsuperscript{1}},\\
  \textbf{Thanathai Lertpetchpun\textsuperscript{1}},
  \textbf{Najim Dehak\textsuperscript{2}},
    \textbf{Dani Byrd\textsuperscript{1}},
  \textbf{Shrikanth Narayanan\textsuperscript{1}}\\
  \textsuperscript{1}Signal Analysis and Interpretation Lab (SAIL), University of Southern California\\
  \textsuperscript{2}Center for Language and Speech Processing, Johns Hopkins University\\
  \texttt{jihwan@usc.edu}
}

\begin{document}
\maketitle


\begin{abstract}
Neural audio codecs are increasingly used for low-bitrate speech coding and serve as tokenizers for speech language models, yet it remains unclear how their codebook hierarchies encode speaker traits and speech characteristics.
We systematically study 17 multi-codebook neural audio codecs across seven traits: sex, age, accent, voice quality, emotion, speech flow, and non-verbal sounds.
For each codec, we reconstruct speech using partial codebooks and use machine listeners to compare trait predictions on each reconstruction.
We measure prediction agreement with reweighted Cohen's $\kappa$, and define \textit{trait depth} as a proxy for the codebook depth by which each trait is encoded and recoverable.
Overall, sex has the shallowest trait depth, followed by speech flow, whereas voice quality and age group require deeper stacks; emotion, accent, and non-verbal sounds fall in between.
Within each trait, speech-flow classes are at relatively similar trait depths, whereas emotion, accent, and non-verbal sounds differ more.
Additionally, we explore how certain codec design choices relate to trait depths.
These findings show that codebook hierarchies are trait-dependent and may not exhibit uniform coarse-to-fine structures, motivating trait-specific codec evaluation and design for future applications.
\end{abstract}

\section{Introduction}
\label{sec:intro}

Neural audio codecs are increasingly used for low-bitrate speech coding and as tokenizers for speech language models. Many employ residual vector quantization (RVQ), in which a sequence of codebooks progressively quantizes the residuals left by preceding codebooks, allowing a single model to operate at multiple bitrates \citep{zeghidour2022soundstream,defossez2023encodec}. Speech language models often exploit this hierarchy by generating the first codebook or first few codebooks separately from the rest \citep{borsos2023audiolm,chen2025valle}. Several recent codecs further assign a distinct role to the first codebook through teacher distillation~\citep{zhang2024speechtokenizer,defossez2024moshi,ye2025xcodec}.


Existing codec evaluations typically assess linguistic content, speaker similarity, and perceptual quality~\citep{wu2024codecsuperb,shi2024espnetcodec,mousavi2026dasb}.
RVQ codebook stacks are commonly described as progressing from coarse to fine-grained features~\citep{borsos2023audiolm, soundstorm, chen2025valle}; however, it remains unclear at what depth specific speaker traits and speech characteristics are preserved along this stacked progression. These traits, including sex, age, accent, voice quality, emotion, speech flow, and non-verbal sounds\footnote{Non-verbal sounds are not speech in the precise definition, but for brevity, we group them with speech characteristics.} (e.g., laughter and coughing), are important for responsive spoken-dialogue systems, accessible speech technologies, and equitable performance across different speakers.

In this work, we investigate how these codecs encode speaker traits and speech characteristics across their codebook hierarchies, going beyond conventional codec evaluation such as speaker similarity and downstream speech-recognition tasks.
We focus on three research questions: \textbf{RQ1:} \emph{By what codebook depth is each trait effectively preserved?}; \textbf{RQ2:} \emph{For each trait, are its classes preserved by similar codebook depths?}; and \textbf{RQ3:} \emph{How do codec design choices affect trait depths?}

To answer these questions, we investigate 17 neural audio codecs. For each codec, we first encode and decode speech utterances using cumulative prefixes of its codebooks. We then use multiple machine listeners for different traits and measure reweighted agreement using Cohen's $\kappa$ to determine whether they make the same decisions on each reconstruction as with the original audio.
We define \textit{trait depth} as the required relative codebook depth where a trait's reweighted $\kappa$ first reaches 90\% of its full-depth value, and use it as a proxy indicator for how each trait is encoded and recoverable in the codebook hierarchy.
We then compare measured trait depths across different neural codecs with speaker traits and speech characteristics.

Our findings suggest that sex has the shallowest trait depth, followed by speech flow. 
Voice quality and age group have deeper trait depths, with the remaining traits and characteristics in between.
Trait depths also differ across classes within a trait.
Speech flow types exhibit similar trait depths, whereas non-verbal sounds differ more in depths, with laughter preserved in earlier codebooks than sounds such as sneezing.
Within matched codec pairs, a higher framerate preserves more information about most traits in earlier codebooks, and streaming has little effect.
Our results suggest a trait-specific account in codec hierarchies, showing that speaker traits and speech characteristics may not in fact follow a uniform coarse-to-fine structure across codebook depths.

\begin{figure*}[t]
  \centering
  \includegraphics[width=\textwidth]{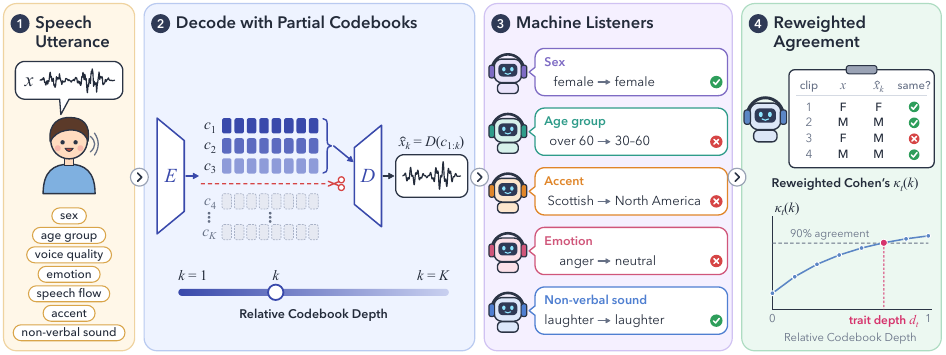}
  \caption{Overview of the protocol.
  (1)~A speech utterance $x$ with different speaker and speech traits.
  (2)~A neural audio codec encodes $x$ into $K$ codebooks and decodes with partial codebooks $\hat{x}_k = D(c_{1:k})$.
  (3)~Machine listeners predict each trait on $x$ and on every $\hat{x}_k$.
  (4)~The reweighted agreement (Cohen's $\kappa$) between $\hat{x}_k$ and $x$ is measured; and \textit{trait depth} $d_t$ is defined as the relative codebook depth where $\kappa$ first reaches 90\% of its full-depth value.}
  \label{fig:overview}
\end{figure*}

\section{Related Work}
\label{sec:related}

\subsection{Neural Audio Codecs}
\label{sec:related-codecs}

Neural audio codecs represent audio using discrete codebooks, often arranged hierarchically. In RVQ, each codebook quantizes the residual left by the preceding codebooks, allowing reconstruction from the first $k$ codebooks at a lower bitrate~\citep{zeghidour2022soundstream,defossez2023encodec,kumar2023dac}.
Other codecs quantize at multiple temporal resolutions or replace RVQ with a single codebook or finite scalar quantization \citep{siuzdak2024snac, xin2024bigcodec,dellalibera2025focalcodec,parker2025taae,casanova2025nanocodec}.

Recent codecs train the first codebook through teacher distillation.
SpeechTokenizer and Mimi distill HuBERT~\cite{hubert} and WavLM~\cite{chen2022wavlm} representations into the first codebook \citep{zhang2024speechtokenizer,defossez2024moshi}, respectively, while
X-Codec, DualCodec, and JHCodec incorporate self-supervised features~\citep{ye2025xcodec,li2025dualcodec,lee2026jhcodec}.
PAST supervises the first codebook with phonemes \citep{hartuv2025past}, and FlexiCodec uses ASR features \citep{li2026flexicodec}.

\subsection{Probing Neural Audio Codecs}
\label{sec:related-probing}

RVQ codebook stacks are commonly described as progressing from coarse to fine~\citep{soundstorm}, yet which information is encoded at each depth remains unclear. Earlier codebooks are generally assumed to capture coarse properties such as speaker identity and recording conditions, while later codebooks recover finer acoustic details \citep{borsos2023audiolm,chen2025valle}.
Teacher distillation places linguistic content in the first codebook and leaves the remaining paralinguistic and timbre information to later codebooks \citep{zhang2024speechtokenizer}. These accounts, however, do not identify where specific speaker traits and speech characteristics appear to be encoded. Furthermore, they are conventionally evaluated only through intelligibility or speaker similarity.

Existing evaluations are also limited to relatively few traits or codecs. Codec benchmarks assess tasks such as speech recognition, speaker verification, and emotion recognition on reconstructed audio \citep{wu2024codecsuperb,shi2024espnetcodec} or using models trained directly on codec tokens \citep{mousavi2026dasb,deng2025codecbench}. Studies of paralinguistic preservation generally examine one trait at a time, such as emotion \citep{ren2024emocodec} or accent \citep{huang2026codecmosaccent}.
Code-level analyses trace content, speaker identity, and pitch in codec representations \citep{sadok2025interpretability}, identify primarily phonetic rather than lexical-semantic structure \citep{shi2026codecprobing}, and locate or manipulate accent using sparse autoencoders \citep{wang2026saeaccent}.
Analyses of Mimi~\citep{defossez2024moshi} further show that its first codebook represents context-dependent phone realizations \citep{saloev2026mimi}, captures segmental information more strongly than prosody \citep{ballier2026mimiprosody}, and nevertheless contains information about English nuclear tunes \citep{vigneaux2026tunes}.
A systematic account of where diverse speaker traits and speech characteristics appear across codebook hierarchies remains lacking.

\subsection{Speaker Traits and Speech Characteristics}
\label{sec:related-traits}

Apart from linguistic content, speech conveys cues to relatively stable speaker traits, such as sex, age, and accent, as well as dynamic properties such as emotion \citep{schuller2013computational}. Spoken-dialogue benchmarks evaluate whether models perceive these traits \citep{ao2024sdeval,yang2024airbench}, while style-prompted text-to-speech systems describe voices using attributes such as nasal, husky, or guttural \citep{diwan2025paraspeechcaps}.

Pretrained trait predictors provide machine listeners for many of these traits. VoxProfile~\citep{feng2025voxprofile} is a comprehensive speaker trait and speech characteristic predictor, including sex, age, accent, voice quality, emotion, and speech flow. 
Voxlect~\citep{feng2026voxlect} covers dialects and regional varieties, while AudioSet-based models such as AST and SSLAM recognize non-verbal sounds including laughter, sighs, and coughs \citep{gemmeke2017audioset,gong2021ast,alex2025sslam,gong2022vocalsound}.
Other examples of trait predictors include CommonAccent and GenAID for accent \citep{zuluagagomez2023commonaccent,zhong2025accentbox}, emotion2vec for emotion and SenseVoice for sound events \citep{ma2024emotion2vec,an2024funaudiollm}, and a wav2vec~2.0 model for age and sex \citep{burkhardt2023agegender}.
We use these pretrained trait predictors as {\em machine listeners}, comparing their decisions on original and reconstructed speech rather than training and exploiting codec-specific probes.

\section{Method}
\label{sec:method}

\subsection{Probing Protocol}
\paragraph{Neural Audio Codecs.}
\label{sec:codecs}
We explore 17 publicly available multi-codebook neural audio codecs spanning various design choices, as summarized in Table~\ref{tab:codecs}. 
Their token framerates range from 12.5 to 75\,Hz; they operate at sampling rates of 16 or 24\,kHz and contain between 4 and 32 codebooks.
The set includes codecs with and without teacher distillation (e.g., from self-supervised speech models) and with and without streaming.

\begin{table*}[t]
  \centering
  \small
  \begin{tabular}{lrrrrcc}
    \toprule
    Codec & Framerate & Sampling Rate & Number of & Bitrate & Distillation & Streaming \\
     & (Hz) & (kHz) & Codebooks & (kbps) & & \\
    \midrule
    EnCodec \citep{defossez2023encodec} & 75 & 24 & 32 & 24 & \xmark & \cmark \\
    DAC \citep{kumar2023dac} & 50 & 16 & 8$^{\dagger}$ & 4 & \xmark & \xmark \\
    FunCodec 50\,Hz \citep{du2024funcodec} & 50 & 16 & 32 & 16 & \xmark & \xmark \\
    FunCodec 25\,Hz & 25 & 16 & 32 & 8 & \xmark & \xmark \\
    SpeechTokenizer \citep{zhang2024speechtokenizer} & 50 & 16 & 8 & 4 & \cmark & \xmark \\
    Mimi \citep{defossez2024moshi} & 12.5 & 24 & 32 & 4.4 & \cmark & \cmark \\
    HILCodec \citep{ahn2024hilcodec} & 75 & 24 & 8 & 6 & \xmark & \cmark \\
    X-Codec \citep{ye2025xcodec} & 50 & 16 & 8 & 4 & \cmark & \xmark \\
    MaskGCT codec \citep{wang2025maskgct} & 50 & 24 & 12 & 6 & \xmark & \xmark \\
    TAAE \citep{parker2025taae} & 25 & 16 & 4 & 0.95 & \xmark & \xmark \\
    DualCodec 12.5\,Hz \citep{li2025dualcodec} & 12.5 & 24 & 8 & 1.23 & \cmark & \xmark \\
    DualCodec 25\,Hz & 25 & 24 & 12 & 3.1 & \cmark & \xmark \\
    PAST \citep{hartuv2025past} & 50 & 16 & 8 & 4 & \cmark$^{\ddagger}$ & \xmark \\
    PAST-streamable & 50 & 16 & 8 & 4 & \cmark$^{\ddagger}$ & \cmark \\
    MOSS \citep{gong2026moss} & 12.5 & 24 & 32 & 4 & \xmark & \cmark \\
    JHCodec \citep{lee2026jhcodec} & 50 & 16 & 8 & 4 & \cmark & \cmark \\
    FlexiCodec \citep{li2026flexicodec} & 12.5 & 16 & 25 & 3.79 & \cmark & \xmark \\
    \bottomrule
    \multicolumn{7}{l}{\tiny $^{\dagger}$DAC at 16\,kHz, with 8 of its 12 codebooks; $^{\ddagger}$ PAST is grouped with distilled codecs by first-codebook supervision.}
  \end{tabular}
  \caption{The 17 neural audio codecs with multiple codebooks and their codec design.}
  \label{tab:codecs}
\end{table*}

\paragraph{Decode with Partial Codebooks.}
\label{sec:resynthesis}
To examine how much of each trait persists at different codebook depths, we reconstruct each utterance using the first $k$ codebooks.
This partial-decoding process is formulated as follows: an input utterance $x$ is first encoded into codes $c_{1:K}=E(x)$, and its reconstruction from the first $k$ codebooks is given by $\hat{x}_k=D(c_{1:k})$, where $E$ and $D$ denote the encoder and decoder, respectively, and $K$ is the total number of codebooks in the codec.

\paragraph{Machine Listeners.}
\label{sec:traits}

To investigate how much speaker traits and speech characteristics are preserved in each partial reconstruction, we feed $\hat{x}_k$ into multiple machine listeners that predict these traits.
As summarized in Table~\ref{tab:traits}, we focus on seven constructs: sex, age group, accent, voice quality, emotion, speech flow, and non-verbal sounds.
We follow the speaker trait and speech characteristic definitions used by VoxProfile \citep{feng2025voxprofile}, Voxlect \citep{feng2026voxlect}, and VocalSound \citep{gong2022vocalsound}, and refer to each category as a \textit{trait} throughout the paper for simplicity.
We use Voxlect and SSLAM as the representative machine listeners for accent and non-verbal sounds, respectively, and VoxProfile for the remaining traits, given their strong prediction performance. Each machine listener outputs a decision for its designated trait on the original utterance $x$ and on reconstruction $\hat{x}_k$.

We also use auxiliary machine listeners to reduce dependence on any specific predictor, when available (Table~\ref{tab:traits}).
The agreement scores produced by the auxiliary listeners strongly align with those of the representative listeners, with Spearman's $\rho$ ranging from 0.87 to 0.98 (Appendix~\ref{app:listeners}).

\begin{table*}[t]
  \centering
  \small
  \setlength{\tabcolsep}{4pt}
  \begin{tabular}{@{}>{\raggedright\arraybackslash}p{1.9cm}>{\raggedright\arraybackslash}p{6.4cm}>{\raggedright\arraybackslash}p{3.7cm}>{\raggedright\arraybackslash}p{3.1cm}@{}}
    \toprule
    Trait & Labels & Machine Listeners & Dataset \\
    \midrule
    Sex\par\vspace{3pt}Age & female, male\par\vspace{3pt}under 30, 30--60, over 60 & VoxProfile; audeering & TIMIT, VoxCeleb, Common Voice \\
    \midrule
    Accent & North America, (England) English, Scottish, Irish, Northern Irish, Welsh, Oceania, South African, Germanic, Romance, Slavic, Semitic, East Asia, Southeast Asia, South Asia, other & Voxlect; GenAID; CommonAccent ECAPA & Voxlect English test set (Section~\ref{sec:datasets}) \\
    \midrule
    Voice quality & \textit{Pitch}: shrill, nasal, deep \newline \textit{Rhythm}: singsong, pitchy, flowing, monotone, staccato, punctuated, enunciated, hesitant \newline \textit{Clarity}: crisp, slurred, lisp, stammering \newline \textit{Texture}: silky, husky, raspy, guttural, vocal fry \newline \textit{Volume}: booming, authoritative, loud, hushed, soft & VoxProfile & ParaSpeechCaps \\
    \midrule
    Emotion & anger, contempt, disgust, fear, happiness, neutral, sadness, surprise, other & VoxProfile; emotion2vec+& IEMOCAP \\
    \midrule
    Speech flow & fluent, block, prolongation, sound repetition, word repetition, interjection & VoxProfile & FluencyBank, SEP-28k \\
    \midrule
    Non-verbal & laughter, sigh, cough, throat clearing, sneeze, sniff & SSLAM; AST; SenseVoice & VocalSound \\
    \bottomrule
  \end{tabular}
  \caption{Speaker traits and speech characteristics used for analysis and their labels, adopted from VoxProfile~\citep{feng2025voxprofile}, Voxlect~\citep{feng2026voxlect}, and VocalSound~\citep{gong2022vocalsound}; machine listeners used to predict each trait; and the corresponding datasets.}
  \label{tab:traits}
\end{table*}

\paragraph{Reweighted Agreement.}
\label{sec:metrics}
We measure agreement between the machine listeners' decisions on $\hat{x}_k$ and $x$. To mitigate potential biases in across-trait and within-trait comparisons from differences in listener accuracy and class distribution, we compute Cohen's $\kappa$ on reweighted decision distributions.

For trait $t$ with decisions $y_i(k)$ on sample $i$ at depth $k$, Cohen's $\kappa$ \citep{cohen1960kappa} is defined as
\begin{equation}
\kappa_t(k) = \frac{p_o - p_e}{1 - p_e},
\label{eq:kappa}
\end{equation}
where $p_o = \frac{1}{N}\sum_i \mathbf{1}[y_i(k)=y_i(0)]$ is the observed agreement over the $N$ samples and $p_e = \sum_c \pi_c(k)\,\pi_c(0)$ is the agreement expected by chance. Here, $\pi_c(k)$ denotes the proportion of decisions assigned to class $c$ at depth $k$, and $k=0$ denotes the decisions on the original utterance $x$ by a machine listner. For voice quality, where each clip may be assigned multiple attributes, $p_o$ is pooled over all attribute decisions, while $p_e$ is computed separately for each attribute and then averaged.

As a proxy indicator of the codebook depth by which each trait is prominently encoded and recoverable, we define \textit{trait depth} $d_t$ as the relative codebook depth at the reweighted $\kappa$ first reaches 90\% of its full-depth value:
\begin{equation}
  d_t = \min\Bigl\{ \frac{k-1}{K-1} : \kappa_t(k) \ge 0.9\,\kappa_t(K) \Bigr\}.
  \label{eq:depth}
\end{equation}
We assume a trait is encoded by depth $k$ when the machine listeners' decisions on $\hat{x}_k$ largely match those on $x$; trait encoding is thus inferred by the decoder and the machine listeners.

We reweight $\kappa$ to mitigate potential biases arising from class prevalence, class balance, and differences in prediction accuracy among machine listeners.
We follow the principles from post-stratification and importance weighting \citep{horvitz1952sampling,fleiss2003rates,shimodaira2000covariate}. Each decision on $x$ is assigned to one of four strata defined by whether the decision is correct or incorrect relative to the label and whether its predicted class is a majority or minority class. Sample $i$ in stratum $s$ receives weight
$w_i=\frac{\pi_s^{\mathrm{ref}}}{\pi_s^t},$
where $\pi^t$ is the stratum distribution for trait $t$ and $\pi^{\mathrm{ref}}$ is the mean distribution across the seven traits.
The reweighted $\kappa$ is obtained by replacing the observed agreement and class proportions with their weighted counterparts.
The weights depend only on the decisions made on $x$ and therefore remain fixed across codecs and codebook depths.

\subsection{Datasets}
\label{sec:datasets}

We primarily use the datasets used to evaluate each representative machine listener (Table~\ref{tab:traits}).
For sex and age group, we use the VoxProfile test sets comprising TIMIT \citep{garofolo1993timit}, VoxCeleb \citep{nagrani2017voxceleb,hechmi2021voxceleb}, and Common Voice \citep{ardila2020commonvoice}.
For accent, we use Voxlect's English test set~\citep{feng2026voxlect}, which includes Common Voice~\citep{ardila2020commonvoice}, CSLU Foreign Accented English~\citep{lander2007cslufae}, EdAcc~\citep{sanabria2023edacc}, the British Isles accent corpora~\citep{demirsahin2020britishisles}, L2-ARCTIC~\citep{zhao2018l2arctic}, TIMIT~\citep{garofolo1993timit}, VoxPopuli~\citep{wang2021voxpopuli}, ESLTTS~\citep{wang2024usat}, Fair-Speech~\citep{veliche2024fairspeech}, ParaSpeechCaps~\citep{diwan2025paraspeechcaps}, Nigerian English~\citep{google2019nigerianenglish}, and Hispanic English~\citep{byrne2014hispanic}.
For voice quality, we use the ParaSpeechCaps test split \citep{diwan2025paraspeechcaps}, which is also used to evaluate VoxProfile. For emotion, we use a randomly selected 10\% subset of IEMOCAP \citep{busso2008iemocap}, which serves as a held-out evaluation set for VoxProfile. For speech flow, we randomly sample 20\% of the test sets from SEP-28k \citep{lea2021sep28k} and FluencyBank \citep{ratner2018fluencybank}. For non-verbal sounds, we use the VocalSound test split \citep{gong2022vocalsound}. More details are in Table~\ref{tab:testsets} in Appendix~\ref{app:listeners}.


\section{Results and Discussion}
\label{sec:results}
\begin{figure*}[t]
  \centering
  \includegraphics[width=\textwidth]{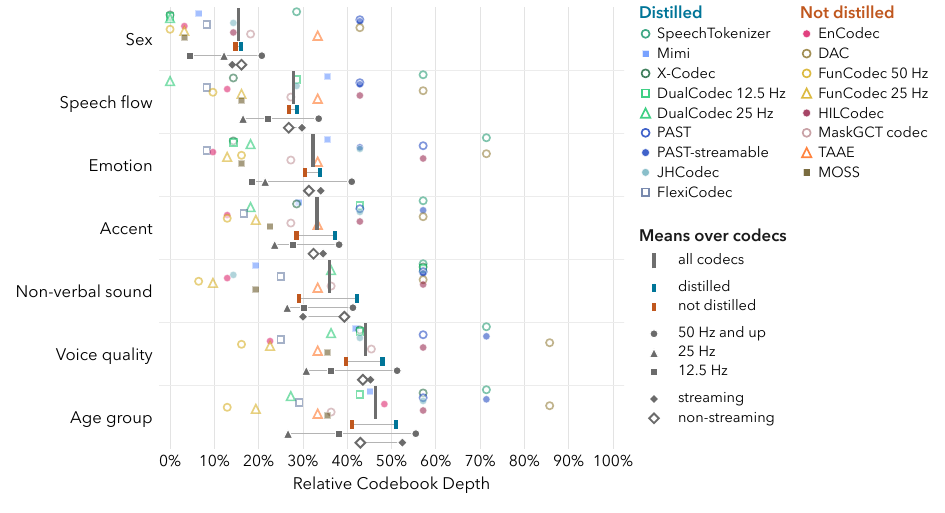}
  \caption{Trait depths $d_t$ for each codec for each trait, the relative codebook depth at which the reweighted $\kappa$ first reaches 90\% of its full-depth value. Each dot is a codec and the gray bars indicate the mean values. Cool and warm colors indicate distilled versus non-distilled codecs, respectively. Different framerates and streamability have different markers.}
  \label{fig:depth}
\end{figure*}

We analyze the results from three main perspectives. First, the across-trait analysis explores the codebook depth by which each trait is encoded and recoverable. 
Second, the within-trait analysis investigates how different classes in a trait exhibit similar or different trait depths.
Third, we examine how codec design choices, including distillation, framerate, and streaming capability, affect trait depths.
We provide an interactive webpage (\href{https://jihwanlee15.github.io/codec-trait-depth/}{\nolinkurl{jihwanlee15.github.io/codec-trait-depth}})
that allows readers to explore the results from multiple perspectives.

\subsection{Across-Trait Analysis}
\label{sec:across}
\begin{figure*}[!t]  
  \centering
  \includegraphics[width=\textwidth]{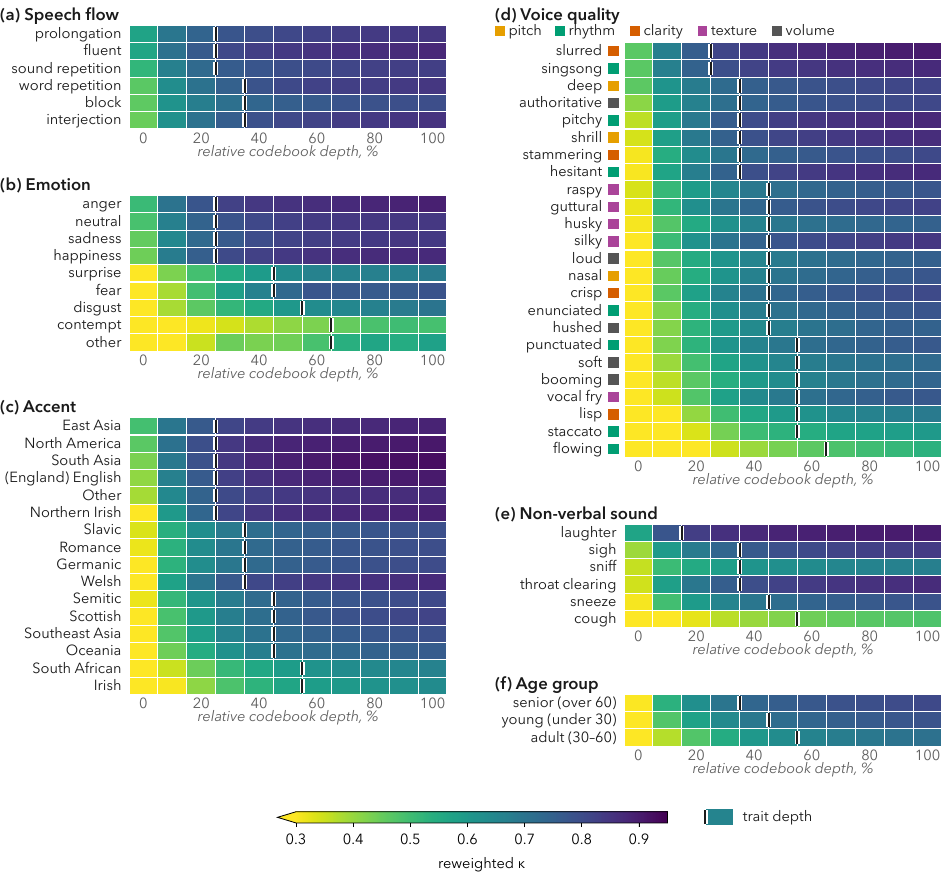}
  \caption{Per-class reweighted $\kappa$ by relative codebook depth, averaged over the 17 codecs. Darker cells indicate higher agreement, and a tick marks \textit{trait depth}, i.e., the first depth where $\kappa$ reaches 90\% of its full-depth value.}
  \label{fig:classes}
\end{figure*}
We first explore trait depths across different traits, by which codebook depth each trait is encoded and recoverable.
Figure~\ref{fig:depth} exhibits trait depth $d_t$ of each codec for each trait $t$, that is, how much relative codebook depth is required for the reweighted $\kappa$ to achieve 90\% of its full-depth value.
Averaging across 17 different neural codecs shows a trend of trait depths for different traits.
Sex appears to show shallower trait depth,
followed by speech flow.
Voice quality and age group have deeper trait depths, and emotion, accent, and non-verbal sounds are in between.

One possible interpretation may relate to the temporal scale over which trait-indicating cues are distributed, whether at the utterance, supra-segmental/syllabic, segmental, or sub-segmental levels.
For instance, speech flow or accent may require longer context at timescales more similar to utterance lengths to identify, whereas voice quality may be characterized from shorter intervals more at the timescale of phone- or syllable level segments.
A second possible interpretation
involves the extent and type of articulatory variation involved in producing the speech signal.
Accent differences may depend more strongly on articulatory patterns that shape the vocalic inventories, for example, whereas voice quality may rely less directly on supralaryngeal articulatory patterning and more specifically on slowly varying source properties. 
Many non-verbal vocalizations may require coordination among supralaryngeal, laryngeal, and respiratory actions at timescales that are quite variable, ranging from variable length laughs and isolated coughs and sniffs, and intermediate timescale and possibly oscillatory throat clearing.
Sex determination from speech may behave in a way more independent from timescale in that it may be relatively easy to encode if cues such as pitch and/or spectral envelope are sufficient for confident recovery distinguishing the two categories available.
Further investigation is needed to verify these hypotheses.

\subsection{Within-Trait Analysis}
\label{sec:within}

We next investigate within-trait variation, whether different classes are encoded and recoverable by similar or different codebook depths.
Figure~\ref{fig:classes} shows the reweighted $\kappa$ of each class by relative codebook depth, averaged over the 17 codecs, with a tick at each class's 90\% threshold (i.e., trait depth).
For most traits, we find different trait depths across classes, each class being encoded and recoverable by different codebook depths.


\paragraph{Speech Flow.}

As shown in Figure~\ref{fig:classes}(a), different classes in speech flow exhibit more similar trait depths than the other traits. Fluency and all disfluency types require a relative codebook depth of approximately 30--40\% to reach 90\% of their full-depth $\kappa$.
This suggests that different speech flow types are encoded and recoverable by similar codebook depths.


\paragraph{Emotion.}
Emotion categories exhibit different trait depths, indicating that they are encoded and recoverable by different depths of codebooks.
The four common emotion categories (i.e., anger, neutral, sadness, and happiness) tend to show shallower trait depths, while less commonly represented emotions, such as disgust, surprise, or fear, appear to exhibit deeper trait depths.
One likely explanation is that common emotion labels are more prevalent in speech data used to train neural codecs and are therefore represented more effectively.

\paragraph{Accent.}

Among English accents, more widely represented English varieties tend to show shallower trait depths, whereas less represented accents, such as Irish and South African English, exhibit deeper trait depths.
Similar to emotion analysis, this may reflect how well these varieties are represented in codec training data, where more represented English varieties are better captured by neural codecs.
A similar pattern has been frequently reported in speech recognition, where systems often perform worse for underrepresented accents~\citep{tatman2017gender,markl2022language,sanabria2023edacc}.

\paragraph{Voice Quality.}

We group the voice-quality characteristics into five dimensions (pitch, rhythm, clarity, texture, and volume) as shown in Figure~\ref{fig:classes}(d). Among these dimensions, texture-related, or timbral, qualities tend to show mid-level trait depths, indicating that they are encoded more in the middle codebooks. Within the pitch dimension, deep and shrill voices appear to have shallower trait depths.
Within the rhythm dimension, intonation-related attributes such as singsong and pitchy styles show shallower trait depths, whereas timing-related attributes such as staccato and flowing appear in deeper trait depths. The clarity and volume dimensions show mixed patterns, with part of the attributes showing shallow trait depths and others being deeper.


\paragraph{Non-verbal Sounds.}



For non-verbal sounds, those that involve some degree of voicing tend to show shallower trait depths, indicating that they are encoded and recoverable by earlier codebooks, while other sounds that are more brief or more related to reflexive respiratory actions appear to be encoded in deeper trait depths.
For instance, laughter, which involves some degree of voicing, requires only 20\% relative codebook depth to reach 90\% of its full-depth $\kappa$, whereas sneezing or coughing, which are more brief, impulsive, and related to respiratory actions, require 50--60\% relative codebook depth.


\paragraph{Age Group.}

All age groups tend to show broadly similar trait depths.
Senior adults over 60 years tend to be in shallower trait depths, followed by young adults aged 18--30 years, whereas adults aged 30--60 years are in slightly deeper trait depths.
This contrasts with the expected training-set distributions of most neural audio codecs, in which adults aged 30--60 years likely constitute the largest age group in the datasets used.



\subsection{Codec Design}
\label{sec:design}


We next investigate how codec designs affect the trait depths.
We focus on three aspects: framerate, distillation, and streaming.
As multiple design factors often vary simultaneously across codecs, where possible, we analyze matched comparisons designed to vary a single attribute.
We also discuss general trends observed in averaging across codecs.

\paragraph{Framerate.}

Within the same codec design, a higher framerate generally preserves more information about most traits.
For instance, DualCodec at 25\,Hz shows shallower trait depth than its 12.5\,Hz version for most traits. Similarly, FunCodec at 50\,Hz retains more information in earlier codebooks than its 25\,Hz version for all seven traits. 
The aggregate pattern across codecs is reversed. Codecs operating at 50\,Hz or higher show deeper trait depths for all seven traits than those operating at lower framerates. 
As noted earlier, this aggregate comparison mixes other design choices, including the number of codebooks and the distillation strategy; therefore, this pattern requires further elucidation.


\paragraph{Distillation.}

We compare the overall effect of teacher distillation on information preserved in different codebook depths. 
For the first codebook, codecs trained with teacher distillation, on average, preserve more information about most traits than those without distillation.
However, codecs trained using teacher distillation show deeper trait depths, indicating that they still require deeper codebook stacks to reach 90\% of their full-depth $\kappa$.
This pattern suggests that distillation improves trait preservation in the first codebook, while later codebooks still provide substantial additional information relative to the full-depth value.
It therefore does not necessarily indicate that trait information is concentrated in the first codebook.
As the groups also differ in the number of codebooks $K$ and other design choices, this aggregate comparison should be interpreted cautiously.

\paragraph{Streamability.}

Streaming makes little difference at the first codebook but has a larger effect at full depth.
In the matched comparison, PAST achieves a higher $\kappa$ than PAST-streamable for four of the seven traits at $k=1$ and for six traits at full depth.
PAST also reaches the 90\% threshold earlier for accent, voice quality, and age group; later for emotion; and exhibits no depth difference for the remaining three traits.
Across all codecs, the mean differences between the streaming and non-streaming groups are small for most traits (Figure~\ref{fig:depth}), except for non-verbal sounds and age groups.


\subsection{Potential Future Applications}
\label{sec:implications}

Our analysis has practical implications for codebook selection, system security, and downstream analysis of codec-based speech models. 
Our analyses of trait and codebook depth profiles can guide which codebooks to retain or regenerate for voice conversion and speaker anonymization.
They may also be useful for revealing manipulation and detection signals, as modifying selected codebook depths could alter or degrade particular traits.
Finally, these profiles indicate how much of a codebook stack downstream analyses may require; tasks involving early-emerging traits may tolerate fewer codebooks, whereas later-emerging traits may require deeper codebook reconstructions.

\section{Conclusion}
\label{sec:conclusion}

We investigate where within their architecture neural audio codecs encode specific speaker traits and speech characteristics along their codebook hierarchies, by analyzing 17 different extant codecs using machine listeners and reweighted agreement on reconstructions generated from partial codebook stacks.
We define \textit{trait depth} as a proxy for the codebook depth by which a trait is encoded and recovered.
We find that each trait emerges at different codebook depths: sex from the early codebooks, voice quality and age group from later, and the others in between.
These differences possibly reflect the temporal scale of the encoded cues underlying each speaker trait or speech characteristic.
Classes within a trait are not encoded uniformly, and 
codec design is also associated with trait depths.
These findings show that any postulated coarse-to-fine hierarchy of neural audio codecs is trait-dependent, motivating codec evaluations and system designs that account for which of these speaker and speech properties are preserved and recoverable by different codebook depths.

\section*{Limitations}

Several limitations exist given our experimental protocol. First, the evaluation aims to measure how closely a machine listener's predictions on a partial reconstruction match its predictions on the original audio, and therefore it does not imply that the reconstruction preserves the trait as perceived by people. For example, a partial reconstruction might retain pitch cues but lose timbre information. As a result, a biological sex classifier from speech could then keep the same prediction as for the original audio, even if human listeners perceive the reconstructed voice differently.

Moreover, decoding can reduce overall audio quality, so changes in predictions may reflect artifacts or machine listener sensitivity to degraded audio rather than the misprediction of a particular trait. As trait depth is defined relative to each codec's full-depth agreement, a codec may reach the 90\% threshold at a shallow depth even when its absolute agreement is low. In the future, human judgments and absolute performance evaluation would help answer these research questions.

The results in this work also draw on different datasets for different speaker traits and speech properties, with substantial differences in their size and speaker demographics. For example, the emotion evaluation consists of 600 clips from 10 speakers, while the voice-quality evaluation contains 13,076 clips from 63 speakers. Differences across speaker traits or speech properties may therefore reflect dataset collection, label reliability, or speaker inclusion apart from codec behavior itself. More importantly, the smaller speaker samples in certain datasets may also make class-level estimates less stable. Future work should consider curating multiple traits and speech properties on a shared, speaker-balanced, and larger-size corpus with consistent annotation and adequate presence of speech properties. 

Additionally, our codec-design comparisons are not fully controlled, as most codecs differ in several design factors simultaneously, limiting causal interpretability. Also, our test sets are predominantly English, so the findings may not generalize to other languages.


\section*{Ethical Considerations}
We analyze specific speaker traits and speech characteristics only to measure their encoding and preservation by codecs, using published classifiers on established datasets and reporting aggregate results. The source labels and datasets have demographic and linguistic limitations. Identifying which codebooks carry particular traits may support anonymization and detection but could also enable trait manipulation; we therefore release no tools for such edits. All codecs, classifiers, and datasets are used under their respective licenses, and no audio is redistributed.

\section*{AI Usage Disclosure}
We used an AI assistant, Claude (Anthropic), for implementation assistance, producing figures and tables, and editing text. The authors designed the study, reviewed these outputs, and take full responsibility for the content of the paper.

\bibliography{references}  

\appendix

\section{Neural Codec Details}
\label{app:codecs}
Table~\ref{tab:prefixes} lists the codebook prefixes $k$ decoded for each codec.
Codecs with up to 12 codebooks are decoded from every prefix, whereas the 25- and 32-codebook codecs are decoded from every prefix up to eight codebooks and on a sparser grid above it.
Every codec reads the same 16\,kHz input, resampled to its input rate, and each $\hat{x}_k$ is resampled back to 16\,kHz.
Table~\ref{tab:checkpoints} lists the checkpoint, code, and license of every codec.

\begin{table}[h]
  \centering
  \small
  \begin{tabular}{lrl}
    \toprule
    Codec & $K$ & Decoded prefixes $k$ \\
    \midrule
    EnCodec & 32 & 1--8, 12, 16, 24, 32 \\
    DAC & 8$^{\dagger}$ & 1--8 \\
    FunCodec 50\,Hz & 32 & 1--8, 12, 16, 24, 32 \\
    FunCodec 25\,Hz & 32 & 1--8, 12, 16, 24, 32 \\
    SpeechTokenizer & 8 & 1--8 \\
    Mimi & 32 &  1--8, 12, 16, 24, 32 \\
    HILCodec & 8 & 1--8 \\
    X-Codec & 8 & 1--8 \\
    MaskGCT codec & 12 & 1--12 \\
    TAAE & 4 & 1--4 \\
    DualCodec 12.5\,Hz & 8 & 1--8 \\
    DualCodec 25\,Hz & 12 & 1--12 \\
    PAST & 8 & 1--8 \\
    PAST-streamable & 8 & 1--8 \\
    MOSS & 32 & 1--8, 12, 16, 24, 32 \\
    JHCodec & 8 & 1--8 \\
    FlexiCodec & 25 & 1--8, 12, 16, 20, 25 \\
    \bottomrule
  \end{tabular}
  \caption{Codebook prefixes decoded for each codec, in the order of Table~\ref{tab:codecs}. $^{\dagger}$The first eight of DAC's 12 codebooks.}
  \label{tab:prefixes}
\end{table}


\section{Machine Listeners}
\label{app:listeners}
The representative and auxiliary machine listeners agree closely (Table~\ref{tab:agreement}). Spearman's $\rho$ between their $\kappa$, as measured, ranges from 0.76 to 0.98 across the 17 codecs at the first codebook ($k = 1$) and from 0.87 to 0.98 over all codec--depth points (All).
We therefore use only the representative machine listeners for the analysis.

\begin{table}[t]
  \centering
  \small
  \setlength{\tabcolsep}{3pt}
  \begin{tabular}{llrr}
    \toprule
    & & \multicolumn{2}{c}{$\rho$ of $\kappa$} \\
    \cmidrule(lr){3-4}
    Trait & Auxiliary Listener & $k = 1$ & All \\
    \midrule
    Sex & audeering & 0.98 & 0.94 \\
    Age group & audeering & 0.95 & 0.96 \\
    Accent & GenAID & 0.79 & 0.97 \\
     & CommonAccent ECAPA & 0.82 & 0.98 \\
    Emotion & emotion2vec+ & 0.82 & 0.87 \\
    Non-verbal sound & AST & 0.79 & 0.95 \\
     & SenseVoice & 0.76 & 0.89 \\
    \bottomrule
  \end{tabular}
  \caption{Spearman's $\rho$ between the $\kappa$ of each trait's representative and auxiliary machine listeners.}
  \label{tab:agreement}
\end{table}

Table~\ref{tab:testsets} lists the number of clips, speakers, and hours for each trait we tested in this work.
Table~\ref{tab:checkpoints-listeners} lists the checkpoint, code, and license of every machine listener.

\begin{table}[t]
  \centering
  \small
  \setlength{\tabcolsep}{3pt}
  \begin{tabular}{@{}lrrr@{}}
    \toprule
    Trait & Clips & Speakers & Hours \\
    \midrule
    Accent & 39,911 & 173 & 82.18 \\
    Sex, age group & 12,619 & 1,295 & 23.45 \\
    Voice quality & 13,076 & 63 & 27.16 \\
    Non-verbal sound & 3,591 & 574 & 4.14 \\
    Emotion & 600 & 10 & 1.00 \\
    Speech flow & 1,622 & 162 & 1.35 \\
    \bottomrule
  \end{tabular}
  \caption{The number of clips, speakers, and total hours of the audio.}
  \label{tab:testsets}
\end{table}

\begin{table*}[t]
  \centering
  \footnotesize
  \setlength{\tabcolsep}{3pt}
  \begin{tabular}{@{}>{\raggedright\arraybackslash}p{2.9cm}>{\raggedright\arraybackslash}p{6.0cm}>{\raggedright\arraybackslash}p{4.0cm}>{\raggedright\arraybackslash}p{2.4cm}@{}}
    \toprule
    Codec & Checkpoint & Code & License \\
    \midrule
    EnCodec & \href{https://huggingface.co/facebook/encodec_24khz}{huggingface.\allowbreak{}co/\allowbreak{}facebook/\allowbreak{}encodec\_\allowbreak{}24khz} & \href{https://github.com/huggingface/transformers}{github.\allowbreak{}com/\allowbreak{}huggingface/\allowbreak{}transformers} & Apache-2.0 \\
    DAC & \href{https://huggingface.co/descript/dac_16khz}{huggingface.\allowbreak{}co/\allowbreak{}descript/\allowbreak{}dac\_\allowbreak{}16khz} & \href{https://github.com/huggingface/transformers}{github.\allowbreak{}com/\allowbreak{}huggingface/\allowbreak{}transformers} & MIT, Apache-2.0 \\
    FunCodec (50, 25\,Hz) & 50\,Hz: \href{https://huggingface.co/alibaba-damo/audio_codec-encodec-en-libritts-16k-nq32ds320-pytorch}{huggingface.\allowbreak{}co/\allowbreak{}alibaba-\allowbreak{}damo/\allowbreak{}audio\_\allowbreak{}codec-\allowbreak{}encodec-\allowbreak{}en-\allowbreak{}libritts-\allowbreak{}16k-\allowbreak{}nq32ds320-\allowbreak{}pytorch}; 25\,Hz: \href{https://huggingface.co/alibaba-damo/audio_codec-encodec-en-libritts-16k-nq32ds640-pytorch}{huggingface.\allowbreak{}co/\allowbreak{}alibaba-\allowbreak{}damo/\allowbreak{}audio\_\allowbreak{}codec-\allowbreak{}encodec-\allowbreak{}en-\allowbreak{}libritts-\allowbreak{}16k-\allowbreak{}nq32ds640-\allowbreak{}pytorch} & \href{https://github.com/modelscope/FunCodec}{github.\allowbreak{}com/\allowbreak{}modelscope/\allowbreak{}FunCodec} & MIT \\
    SpeechTokenizer & \href{https://huggingface.co/fnlp/SpeechTokenizer}{huggingface.\allowbreak{}co/\allowbreak{}fnlp/\allowbreak{}SpeechTokenizer} (\texttt{speechtokenizer\_\allowbreak{}hubert\_\allowbreak{}avg}) & \href{https://github.com/ZhangXInFD/SpeechTokenizer}{github.\allowbreak{}com/\allowbreak{}ZhangXInFD/\allowbreak{}SpeechTokenizer} & Apache-2.0 \\
    Mimi & \href{https://huggingface.co/kyutai/mimi}{huggingface.\allowbreak{}co/\allowbreak{}kyutai/\allowbreak{}mimi} & \href{https://github.com/huggingface/transformers}{github.\allowbreak{}com/\allowbreak{}huggingface/\allowbreak{}transformers} & CC BY 4.0, Apache-2.0 \\
    HILCodec & \href{https://github.com/aask1357/hilcodec/tree/master/onnx}{github.\allowbreak{}com/\allowbreak{}aask1357/\allowbreak{}hilcodec/\allowbreak{}tree/\allowbreak{}master/\allowbreak{}onnx} (\texttt{hil\_\allowbreak{}speech}) & \href{https://github.com/aask1357/hilcodec}{github.\allowbreak{}com/\allowbreak{}aask1357/\allowbreak{}hilcodec} & MIT \\
    X-Codec & \href{https://huggingface.co/hf-audio/xcodec-hubert-librispeech}{huggingface.\allowbreak{}co/\allowbreak{}hf-\allowbreak{}audio/\allowbreak{}xcodec-\allowbreak{}hubert-\allowbreak{}librispeech} & \href{https://github.com/huggingface/transformers}{github.\allowbreak{}com/\allowbreak{}huggingface/\allowbreak{}transformers} & CC BY 4.0, Apache-2.0 \\
    MaskGCT codec & \href{https://huggingface.co/amphion/MaskGCT}{huggingface.\allowbreak{}co/\allowbreak{}amphion/\allowbreak{}MaskGCT} (\texttt{acoustic\_\allowbreak{}codec/}) & \href{https://github.com/open-mmlab/Amphion}{github.\allowbreak{}com/\allowbreak{}open-\allowbreak{}mmlab/\allowbreak{}Amphion} & CC BY-NC 4.0, MIT \\
    TAAE & \href{https://huggingface.co/stabilityai/stable-codec-speech-16k}{huggingface.\allowbreak{}co/\allowbreak{}stabilityai/\allowbreak{}stable-\allowbreak{}codec-\allowbreak{}speech-\allowbreak{}16k} & \href{https://github.com/Stability-AI/stable-codec}{github.\allowbreak{}com/\allowbreak{}Stability-\allowbreak{}AI/\allowbreak{}stable-\allowbreak{}codec} & Stability AI Community License, MIT \\
    DualCodec (12.5, 25\,Hz) & \href{https://huggingface.co/amphion/dualcodec}{huggingface.\allowbreak{}co/\allowbreak{}amphion/\allowbreak{}dualcodec} (12.5\,Hz: \texttt{dualcodec\_\allowbreak{}12hz\_\allowbreak{}16384\_\allowbreak{}4096.safetensors}; 25\,Hz: \texttt{dualcodec\_\allowbreak{}25hz\_\allowbreak{}16384\_\allowbreak{}1024.safetensors}) & \href{https://github.com/jiaqili3/DualCodec}{github.\allowbreak{}com/\allowbreak{}jiaqili3/\allowbreak{}DualCodec} & MIT \\
    PAST, PAST-streamable & \href{https://huggingface.co/slprl/PAST}{huggingface.\allowbreak{}co/\allowbreak{}slprl/\allowbreak{}PAST} (PAST: \texttt{PAST.th}; streamable: \texttt{PAST\_\allowbreak{}streamable.th}) & \href{https://github.com/slp-rl/PAST}{github.\allowbreak{}com/\allowbreak{}slp-\allowbreak{}rl/\allowbreak{}PAST} & MIT \\
    MOSS & \href{https://huggingface.co/OpenMOSS-Team/MOSS-Audio-Tokenizer}{huggingface.\allowbreak{}co/\allowbreak{}OpenMOSS-\allowbreak{}Team/\allowbreak{}MOSS-\allowbreak{}Audio-\allowbreak{}Tokenizer} & \href{https://github.com/OpenMOSS/MOSS-Audio-Tokenizer}{github.\allowbreak{}com/\allowbreak{}OpenMOSS/\allowbreak{}MOSS-\allowbreak{}Audio-\allowbreak{}Tokenizer} & Apache-2.0 \\
    JHCodec & \href{https://huggingface.co/jhcodec/jhcodec_1.4m}{huggingface.\allowbreak{}co/\allowbreak{}jhcodec/\allowbreak{}jhcodec\_\allowbreak{}1.\allowbreak{}4m} (\texttt{jhcodec\_\allowbreak{}mimi\_\allowbreak{}1400000.pt}) & \href{https://github.com/jhcodec843/jhcodec}{github.\allowbreak{}com/\allowbreak{}jhcodec843/\allowbreak{}jhcodec} & MIT \\
    FlexiCodec & \href{https://huggingface.co/jiaqili3/flexicodec}{huggingface.\allowbreak{}co/\allowbreak{}jiaqili3/\allowbreak{}flexicodec} (\texttt{12hz\_\allowbreak{}v1\_\allowbreak{}half}) & \href{https://github.com/AmphionTeam/FlexiCodec}{github.\allowbreak{}com/\allowbreak{}AmphionTeam/\allowbreak{}FlexiCodec} & MIT\\
    \bottomrule
  \end{tabular}
  \caption{Checkpoint, code, and license of the codecs.}
  \label{tab:checkpoints}
\end{table*}

\begin{table*}[t]
  \centering
  \footnotesize
  \setlength{\tabcolsep}{3pt}
  \begin{tabular}{@{}>{\raggedright\arraybackslash}p{2.9cm}>{\raggedright\arraybackslash}p{6.0cm}>{\raggedright\arraybackslash}p{4.0cm}>{\raggedright\arraybackslash}p{2.4cm}@{}}
    \toprule
    Listener & Checkpoint & Code & License \\
    \midrule
    VoxProfile & \href{https://huggingface.co/tiantiaf}{huggingface.\allowbreak{}co/\allowbreak{}tiantiaf}& \href{https://github.com/tiantiaf0627/vox-profile-release}{github.\allowbreak{}com/\allowbreak{}tiantiaf0627/\allowbreak{}vox-\allowbreak{}profile-\allowbreak{}release} & OpenRAIL, RAIL-S 1.1 \\
    Voxlect& \href{https://huggingface.co/tiantiaf/voxlect-english-dialect-whisper-large-v3}{huggingface.\allowbreak{}co/\allowbreak{}tiantiaf/\allowbreak{}voxlect-\allowbreak{}english-\allowbreak{}dialect-\allowbreak{}whisper-\allowbreak{}large-\allowbreak{}v3} & \href{https://github.com/tiantiaf0627/voxlect}{github.\allowbreak{}com/\allowbreak{}tiantiaf0627/\allowbreak{}voxlect} & OpenRAIL, RAIL-S 1.1 \\
    SSLAM & \href{https://huggingface.co/ta012/SSLAM_AS2M_Finetuned}{huggingface.\allowbreak{}co/\allowbreak{}ta012/\allowbreak{}SSLAM\_\allowbreak{}AS2M\_\allowbreak{}Finetuned} (rev.\ \texttt{4500c5d}) & \href{https://github.com/ta012/SSLAM}{github.\allowbreak{}com/\allowbreak{}ta012/\allowbreak{}SSLAM} & MIT \\
    audeering & \href{https://huggingface.co/audeering/wav2vec2-large-robust-24-ft-age-gender}{huggingface.\allowbreak{}co/\allowbreak{}audeering/\allowbreak{}wav2vec2-\allowbreak{}large-\allowbreak{}robust-\allowbreak{}24-\allowbreak{}ft-\allowbreak{}age-\allowbreak{}gender} & \href{https://github.com/huggingface/transformers}{github.\allowbreak{}com/\allowbreak{}huggingface/\allowbreak{}transformers} & CC BY-NC-SA 4.0, Apache-2.0 \\
    GenAID & \href{https://drive.google.com/file/d/1slGrpZSu5g-nF7R-QMCmtGcjN3kw7lQj/view}{drive.\allowbreak{}google.\allowbreak{}com/\allowbreak{}file/\allowbreak{}d/\allowbreak{}1slGrpZSu5g-\allowbreak{}nF7R-\allowbreak{}QMCmtGcjN3kw7lQj/\allowbreak{}view} (\texttt{GenAID\_\allowbreak{}v6}) & \href{https://github.com/jzmzhong/GenAID}{github.\allowbreak{}com/\allowbreak{}jzmzhong/\allowbreak{}GenAID} & Apache-2.0 \\
    CommonAccent ECAPA & \href{https://huggingface.co/Jzuluaga/accent-id-commonaccent_ecapa}{huggingface.\allowbreak{}co/\allowbreak{}Jzuluaga/\allowbreak{}accent-\allowbreak{}id-\allowbreak{}commonaccent\_\allowbreak{}ecapa} & \href{https://github.com/speechbrain/speechbrain}{github.\allowbreak{}com/\allowbreak{}speechbrain/\allowbreak{}speechbrain} & MIT, Apache-2.0 \\
    emotion2vec+ large & \href{https://huggingface.co/emotion2vec/emotion2vec_plus_large}{huggingface.\allowbreak{}co/\allowbreak{}emotion2vec/\allowbreak{}emotion2vec\_\allowbreak{}plus\_\allowbreak{}large} & \href{https://github.com/modelscope/FunASR}{github.\allowbreak{}com/\allowbreak{}modelscope/\allowbreak{}FunASR} & FunASR Model License, MIT \\
    AST & \href{https://huggingface.co/MIT/ast-finetuned-audioset-10-10-0.4593}{huggingface.\allowbreak{}co/\allowbreak{}MIT/\allowbreak{}ast-\allowbreak{}finetuned-\allowbreak{}audioset-\allowbreak{}10-\allowbreak{}10-\allowbreak{}0.\allowbreak{}4593} & \href{https://github.com/huggingface/transformers}{github.\allowbreak{}com/\allowbreak{}huggingface/\allowbreak{}transformers} & BSD-3-Clause, Apache-2.0 \\
    SenseVoice-Small & \href{https://huggingface.co/FunAudioLLM/SenseVoiceSmall}{huggingface.\allowbreak{}co/\allowbreak{}FunAudioLLM/\allowbreak{}SenseVoiceSmall} & \href{https://github.com/modelscope/FunASR}{github.\allowbreak{}com/\allowbreak{}modelscope/\allowbreak{}FunASR} & FunASR Model License 1.1, MIT \\
    \bottomrule
  \end{tabular}
  \caption{Checkpoint, code, and license of machine listeners.}
  \label{tab:checkpoints-listeners}
\end{table*}

\section{Compute Resources}
\label{app:compute}
No model training is done, and all computation is inference and scoring with the released codec checkpoints and machine listeners.
The approximate upper bound of total compute time is about 1,400 GPU hours on NVIDIA A40, L40S, P100, and A100 GPUs. Several workers shared one GPU, as actual compute time may differ.

\section{Additional Results}
\label{app:percodec}

\begin{figure*}[t]
  \centering
  \includegraphics[width=\textwidth]{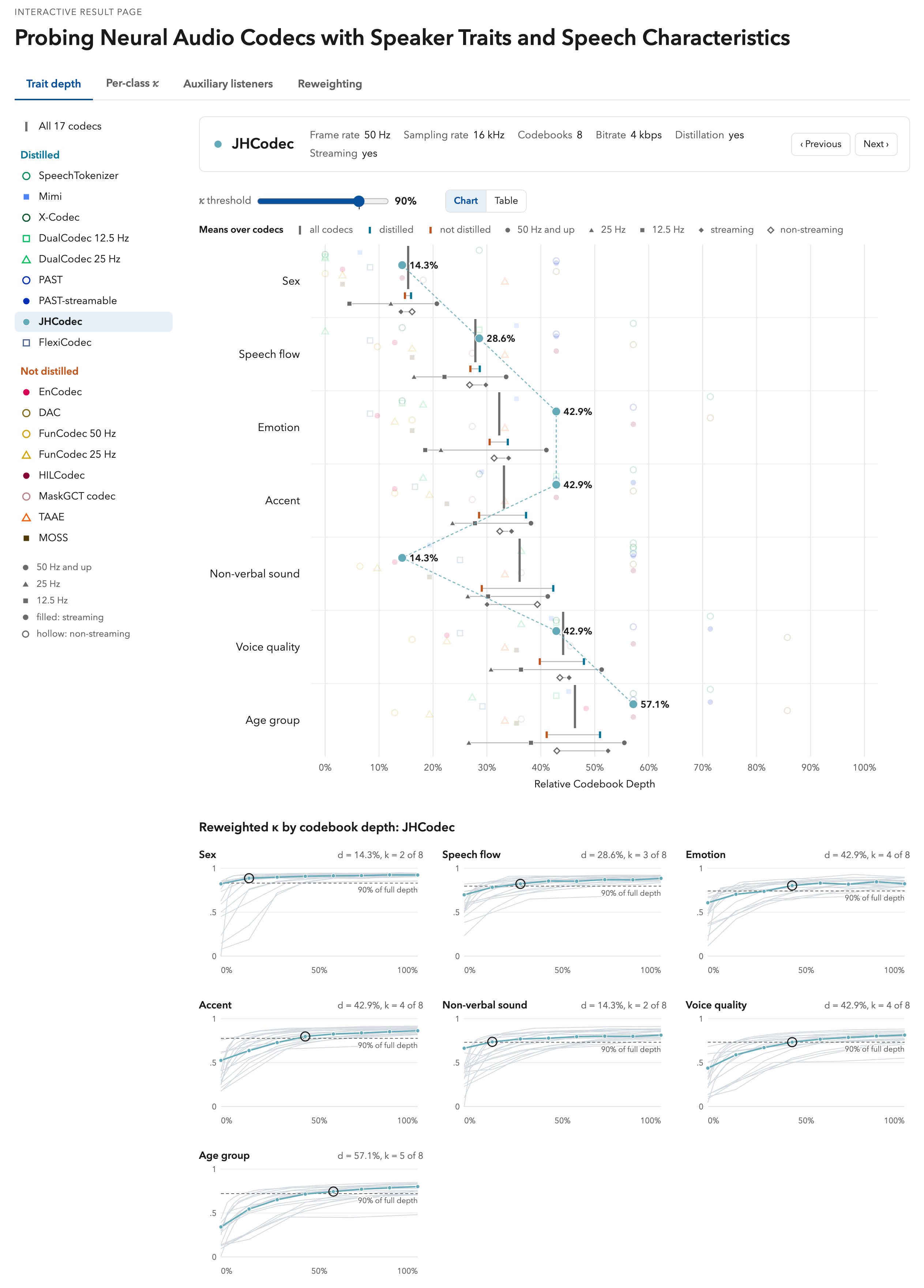}
  \caption{An example screenshot of the interactive webpage.}
  \label{fig:explorer}
\end{figure*}

More results are available on an interactive webpage. On the page, readers can adjust the $\kappa$ threshold that defines trait depth, select individual codecs to see their trait depths and their reweighted $\kappa$ at every decoded prefix, inspect per-class $\kappa$ codec by codec, compare the representative and auxiliary listeners, and see how reweighting changes trait depth relative to listener accuracy and class balance. We highly encourage readers to explore the interactive page: \href{https://jihwanlee15.github.io/codec-trait-depth/}{\nolinkurl{jihwanlee15.github.io/codec-trait-depth}}. Figure~\ref{fig:explorer} shows one of its views.


\end{document}